\documentclass[sigconf]{acmart}

\copyrightyear{2022} 
\acmYear{2022} 
\setcopyright{acmcopyright}\acmConference[WWW '22]{Proceedings of the ACM Web Conference 2022}{April 25--29, 2022}{Virtual Event, Lyon, France}
\acmBooktitle{Proceedings of the ACM Web Conference 2022 (WWW '22), April 25--29, 2022, Virtual Event, Lyon, France}
\acmPrice{15.00}
\acmDOI{10.1145/3485447.3511993}
\acmISBN{978-1-4503-9096-5/22/04}

\AtBeginDocument{%
  \providecommand\BibTeX{{%
    \normalfont B\kern-0.5em{\scshape i\kern-0.25em b}\kern-0.8em\TeX}}}

\usepackage{booktabs}
\usepackage{multirow}
\usepackage{graphicx}
\usepackage{float}

\begin{document}

\title{UniParser: A Unified Log Parser for Heterogeneous Log Data}

\author{Yudong Liu$^*$, Xu Zhang$^*$, Shilin He$^*$, Hongyu Zhang$^\star$, Liqun Li$^*$, Yu  Kang$^*$, Yong Xu$^*$,  Minghua Ma$^*$, Qingwei Lin$^{*}$, Yingnong Dang$^\mathsection$, Saravan Rajmohan$^\dagger$, Dongmei Zhang$^*$}
\affiliation{%
 \institution{Microsoft Research$^*$, Microsoft 365 $^\dagger$,  Microsoft Azure$^\mathsection$,  The University of Newcastle$^\star$}
 \country{}
}

\email{
{yudongliu,xuzhang2,shilin.he,liqli,kay,yox,minghuama,qlin,yidang,saravar,dongmeiz}@microsoft.com
}
\email{hongyu.zhang@newcastle.edu.au}

\renewcommand{\shortauthors}{Y.Liu, X.Zhang, S.He, H.Zhang, L.Li, Y.Kang, Y.Xu, M.Ma, Q.Lin, Y.Dang, S.Rajmohan, D.Zhang}

\begin{abstract}

Logs provide first-hand information for engineers to diagnose failures in large-scale online service systems.
Log parsing, which transforms semi-structured raw log messages into structured data, is a prerequisite of automated log analysis such as log-based anomaly detection and diagnosis.
Almost all existing log parsers follow the general idea of extracting the common part as templates and the dynamic part as parameters.
However, these log parsing methods, often neglect the semantic meaning of log messages. 
Furthermore, high diversity among various log sources also poses an obstacle in the generalization of log parsing across different systems. 
In this paper, we propose UniParser to capture the common logging behaviours from heterogeneous log data.
UniParser utilizes a Token Encoder module and a Context Encoder module to learn the patterns from the log token and its neighbouring context.
A Context Similarity module is specially designed to model the commonalities of learned patterns.
We have performed extensive experiments on 16 public log datasets
and our results show that UniParser outperforms state-of-the-art log parsers by a large margin.
\footnote{Qingwei Lin is the corresponding author of this paper.}

\end{abstract}



\maketitle

{\fontsize{8pt}{8pt} \selectfont
\textbf{ACM Reference Format:}\\
Yudong Liu, Xu Zhang, Shilin He, Hongyu Zhang,Liqun Li,
Yu Kang, Yong Xu, Minghua Ma, Qingwei Lin, Yingnong Dang,
Saravan Rajmohan, Dongmei Zhang. 2022. UniParser: A Unified Log
Parser for Heterogeneous Log Data. In  \textit{Proceedings of the ACM Web Conf. 2022 (WWW’22), April 25--29, 2022, Virtual Event, Lyon, France.} ACM, New York, NY, USA, 9 pages. https://doi.org/10.1145/3485447.3511993}

\section{Introduction}

Online services have surged into popularity in recent years and serve millions of customers on a 24/7 basis, such as Google Search, Bing, Facebook and Twitter.
Although enormous amounts of effort have been resorted to maintain the reliability and availability of these services, in practice, various hardware or software failures are still inevitable, leading to unplanned interruptions of the services. 
Once a failure bursts, operators and developers tend to inspect console logs that record system events and runtime status, to investigate, mitigate and resolve the failure timely.

However, facing the rapid growth volume of raw log messages, it is becoming more and more challenging to identify the valuable information from the enormous log data, even for those experienced engineers~\cite{onion}.
To tackle this problem, \textit{automated log analysis} has emerged in recent years, aiming to automatically analyze the log data with machine learning (ML) or deep learning (DL) techniques. Typical log analysis scenarios consist of log-based anomaly detection~\cite{xu2009detecting, du2017deeplog, zhang2019robust}, diagnosis~\cite{log3c, onion}, failure prediction~\cite{zhang2018prefix, luo2021ntam}, and performance modeling~\cite{chow2014mystery}. Among them, an important and widely-adopted first step is \textit{log parsing},  which parses the \textit{semi-structured} console logs into a \textit{structured} format. After log parsing, the structurized log data are fed into various ML or DL models (e.g., PCA~\cite{xu2009detecting},  LSTM~\cite{deeplog}) for further analysis.

\begin{figure}[t!]
\centering
\includegraphics[width=1.0\linewidth]{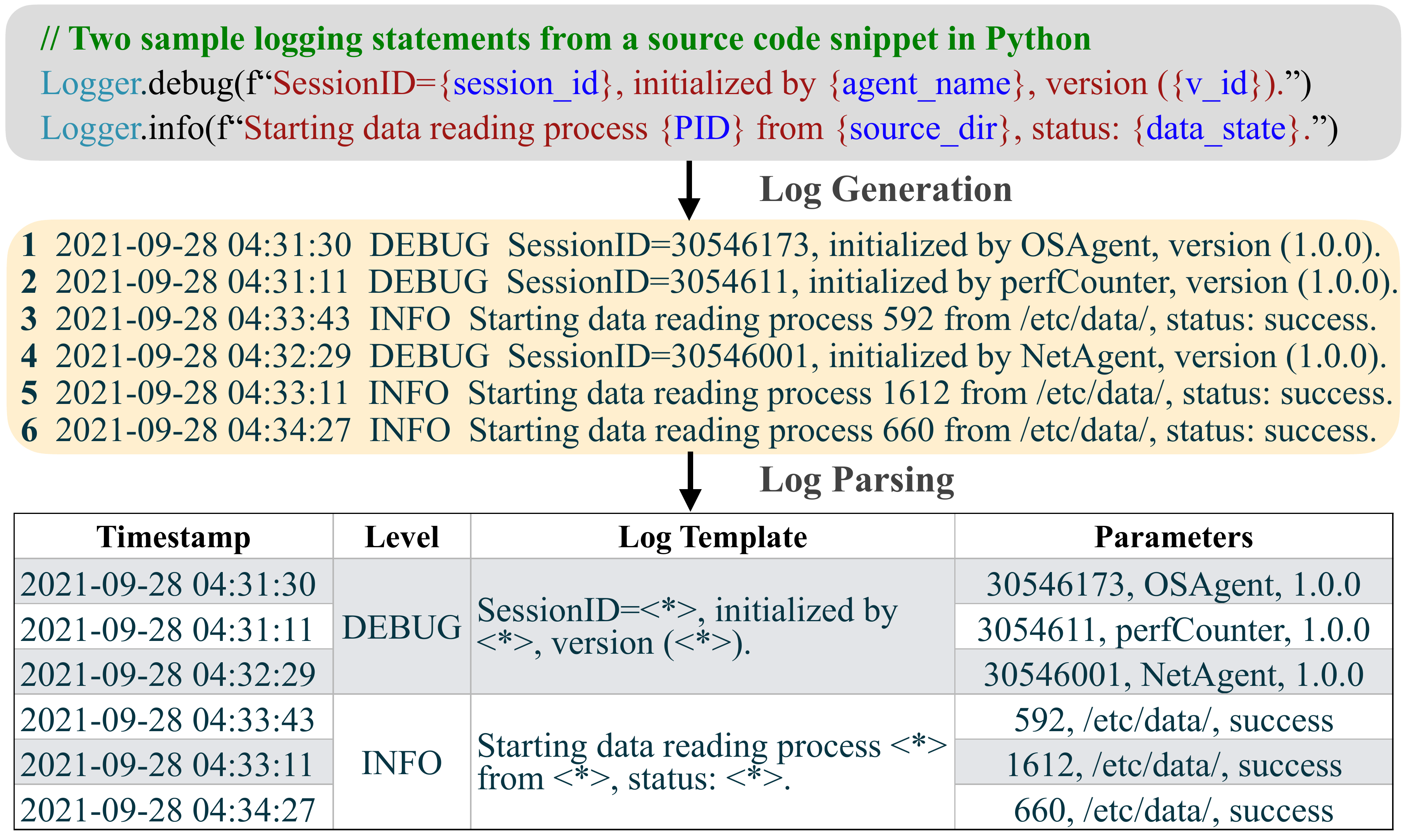}
\caption{An example of log parsing after log generation.}
\label{fig:example}
\end{figure}

As shown in Fig.\ref{fig:example}, logs are generated from various logging statements in the source code during the software execution and collected in an interleaving manner. A typical log contains the log header (e.g., time, level) and the log message, which further consists of two elements: 1) \textit{static} descriptive words fixed in source code to represent the system events, namely \textit{log template}. For example, ``\textit{SessionID=<*>, initialized by <*>, version (<*>).}''; 2) \textit{dynamic} variables, which reflect system runtime status that varies with different executions, called \textit{log parameters}. 
For instance, ``30546173'', ``OSAgent'', and ``1.0.0'' are  parameters generated from three different variables. 
The goal of log parsing is to extract the static log template and the dynamic log parameters from a raw log message.

To enable the log parsing, a straightforward way is to match logs with the source code. However, it is not applicable in practice since the method is ad-hoc and label-intensive when dealing with different logging formats. Besides, the source code is often unavailable, especially for those third-party libraries~\cite{tools}. An alternative approach is manually designing regular expressions based on the generated log data, but it suffers from the low scalability and low accuracy problems~\cite{tools}. To overcome these problems, some data-driven log parsers~\cite{drain, logram, logsig, logcluster} have been proposed in recent years. These approaches follow the same basic paradigm: tokens (e.g., ``SessionID='', ``Starting'') that do not vary with the log messages are templates while the opposite ones (e.g., ``592'', ``1612'', ``660'') are log parameters.  Specifically, a variety of statistical analysis (e.g., prefix tree~\cite{drain}) and data mining techniques (e.g., frequent pattern mining~\cite{logram} and clustering~\cite{logsig}) are leveraged to extract the common parts from raw log messages.

Although making progresses, existing log parsers are still criticized for the unsatisfactory parsing accuracy, which may cause significant adverse effect on follow-up tasks such as log-based anomaly detection~\cite{he2016evaluation}. We summarize two major problems that lead to erroneous parsing results: 1) Existing log parsers only rely on extracting common parts as templates but ignore the semantic meaning of logs. A typical case is illustrated in Fig.\ref{fig:example}.
Considering the semantic meaning of the log message, the directory address "/etc/data" and the return status "success" should apparently be identified as parameters. Here, "/etc/data" is the default configured data reading directory and "success" represents the normal system behaviour, both of which would not be changed often in log messages. Thus, without considering the semantics, existing log parsers tend to mistakenly treat both parameters as the template. 2) Log contents from different services and systems possess high diversity.
It can often be found that the system events as well as the words used in logs produced by different data sources are very different.
As an example, we found that only about 0.5\% of log tokens are shared among log sources generated by three different OS platforms (Linux, Windows and Mac). Thus, it hinders the generality of log parsing across heterogeneous log data sources. 
Facing new log sources, we have to re-accumulate sufficient raw log data as the training materials, adjust hyper-parameters to fit the new log sources, reconfigure the regular expressions for preprocessing, or even build a brand new log parser.

To deal with the above-mentioned problems, in this paper, we point out an important but particular characteristic of log data, i.e, there exist common semantic patterns to indicate templates or parameters regardless of diverse log contents. It is because developers tend to follow the common logging practice for readability. 
For instance, developers habitually nest the parameters in a bracket (e.g., "(1.0.0)") or conventionally set the return status (e.g., "success") as a parameter.
In order to capture this kind of patterns prevailing throughout most log data sources, we propose a unified log parser trained across multiple log sources, named UniParser.
It consists of three modules. Token Encoder module and Context Encoder module are responsible for learning the semantic patterns from the log token itself and its neighbouring context.
Context Similarity module forces our model to focus more on the commonalities of the learned patterns.
After training, UniParser can be directly applied to new log data sources.

We have evaluated UniParser on 16 public log datasets~\cite{log_parsing_dataset}.
UniParser outperforms the state-of-the-art parsers by 12\% on Group Accuracy~\cite{tools} and about 40\% on Message-Level Accuracy. 
More than that, UniParser also can parse millions of logs in only 2 $\sim$ 3 minutes, which is only about half the running time of the most efficient existing parser.

To summarize, our main contributions are as follows:
\vspace{-6pt}
\begin{itemize}
\item We propose UniParser, a unified log parser that can capture the common patterns of templates or parameters across heterogeneous log sources.

\item We evaluated our proposed UniParser on 16 benchmark log datasets and the results show that UniParser performs much better than all existing approaches. Its efficiency and module effectiveness are also verified through the experiments.

\end{itemize}

\section{Background and Motivation}
\label{sec:background}

\subsection{Log Parsing}
Log parsing has been extensively studied in literature~\cite{tools, drain, IPLoM, logsig}.
In general, the ultimate goal of log parsing is to extract the static log template part and the dynamic log parameter part from a raw log message. Developers write their free-text logs, which are composed of some static descriptive words (log template), and some dynamic variables that will be filled up in the log template during the software running period (log parameters). "2021-09-28 04:31:30  DEBUG" is the log header, which is generally easy to extract because its format is fixed in a specific log source.
The "SessionID=<*>, initialized by <*>, version(<*>)" is the log template and "30546173",  "OSAgent" and "1.0.0" are the log parameters filled up into the template placeholder for forming the complete log message.

\subsection{Existing Log Parsers}

\subsubsection{General Idea}
Almost all existing log parsers follow the core idea as below: \textit{extracting common part of raw log messages as log templates and remaining parts are treated as log parameters.}
For example as shown in Fig.\ref{fig:example}, we can first group a series of simliar logs together (shown in brown color) and find that there are some log tokens appear frequently and widely throughout these logs, such as "SessionID", "initialized", "by" in line 1, 2, 4.
These log tokens thus should be identified as the log template part.
On the contrary, other tokens varying in different log messages are regarded as log parameters, such as "OSAgent", "NetAgent" and "PerfCounter".

\subsubsection{Related Work}
\label{ssec:related_work}
In this section,  we introduce how existing log parsers implement the above core idea. These parsers can be categorized by the techniques they adopt:

\paragraph{Frequent Pattern Mining}
Intuitively, the common part of logs should emerge frequently in the whole log dataset.
Therefore, frequent pattern mining technologies were applied widely in the log parsing task.
Typical approaches include SLCT~\cite{SLCT}, LFA~\cite{LFA}, LogCluster~\cite{logcluster} and Logram~\cite{logram}.
These methods firstly traverse over the log data and build frequent itemsets based on tokens, token-position pairs or token n-grams. Then, given the frequent itemset, log messages can be grouped into several clusters and log templates can be extracted from the clusters. SLCT is the first work that applies frequent pattern mining to log parsing~\cite{tools}.
LFA further took the token frequency distribution into account.
LogCluster considered the token position during the process of frequent items mining. 
Logram aims to extract those frequent 2-gram and 3-gram of log tokens, instead of based on single token.

\paragraph{Clustering}
For capturing the common part of logs, another idea is to cluster the similar logs together and identify the common tokens shared within each cluster as its template.
Compared with frequent pattern mining methods, this kind of approaches enables the common part extraction process on the local cluster of similar logs instead of the global log dataset.
LKE~\cite{LKE}, LogSig~\cite{logsig}, LogMine~\cite{Logmine}, SHISO ~\cite{SHISO}, and LenMa~\cite{LenMa} adopt this technology pathway.
Specifically, LKE and LogMine utilize the hierarchical clustering
algorithm to group similar logs based on weighted edit distances.
Instead of conducting clustering on raw log messages directly, LogSig extracts the signature of logs first, on which the clustering is performed then.
SHISO and LenMa are both online log parsing methods, which means they are capable of processing log messages one by one in a streaming manner, which is more practical in real-world scenarios~\cite{tools}.
Both log parsers use the idea of incremental clustering technology.

\paragraph{Heuristics}
Different from general text data, log messages
have some unique characteristics. As such, some log parsers leveraged them to extract common parts as templates.
AEL~\cite{jiang2008abstracting} separates log messages into multiple groups by comparing the occurrences between constant tokens and variable tokens.
IPLoM~\cite{IPLoM} employs an iterative partitioning strategy, which partitions log messages into groups by message length, token position and mapping relation.
Drain~\cite{drain} borrows the idea from prefix tree. 
It builds a fixed-depth tree structure to represent log messages and extracts common templates efficiently.
Spell~\cite{spell} utilizes the longest common subsequence algorithm to parse logs in a stream manner.

\subsection{Limitations of Existing Work}

Although the existing log parsers have achieved good performance in some public datasets, these approaches are still hardly applied in real-world scenarios.
The reason behind lies in their unsatisfying parsing accuracy on complicated real-world log data.
We point out three major problems causing inaccurate parsing results, which could fundamentally challenge the current log parsing paradigm.

\subsubsection{Ignoring semantic meaning of log tokens}
Existing log parsers only consider the static and dynamic properties but neglect the semantic meaning of logs. 
As a consequence, it often gives rise to unreasonable misidentification of log parameters and templates.
At first glance in the example shown in Fig.\ref{fig:example}, some log tokens (such as "success" and "/etc/data/") seem to belong to a part of log template because they do not vary in different log messages.
However, if we take their semantic meaning into consideration, these tokens should be classified as parameters.
For example, "from" is always followed by parameters, such as directory address "/etc/data/" in this illustrated case. Therefore, they are likely to be misidentified as a part of log template. To deal with this problem, most offline log parsers (only support batch processing and all log data are required to be available before parsing)~\cite{tools} have to accumulate sufficient large volume of log messages to guarantee the parameters parts exhibit significant dynamic characteristics.

\subsubsection{Barrier among different log sources}
\label{subsec:log_diversity}
Log contents from different systems or services are characterized by their high diversity. 
As a real example, we performed statistical analysis on collected syslogs from three commonly-used OS platforms\footnote{\url{https://github.com/logpai/loghub}}, i.e., Windows\footnote{collected from Component Based Servicing (CBS) logs}, Mac and Linux.\footnote{both are collected from /var/log/system.log}
We found that they shared only about 0.5\% of common log tokens.
For example, "session opened for user news by (uid=0)" (from Linux log) and "Session: 30546173\_4261722401 initialized by client WindowsUpdateAgent." (from Windows log) are totally different log events and only one word "session" are shared between the two.
High diversity among various log sources poses an obstacle in log parsers generalization across different systems or services. 
Every time when we apply the existing approaches to a new log data source, we have to re-accumulate sufficient raw log data as the training materials, adjust hyper-parameters to adapt to the new log sources, and reconfigure the regular expressions for preprocessing. Especially, for online log parsers (building up a model in advance and process log messages one
by one in a streaming manner~\cite{tools}), we even need to develop a new log parser from scratch.

\subsubsection{Improper evaluation metrics}
Log templates and parameters identification should be treated equally without discrimination. 
For example, on the one hand, most log anomaly detection models, such as LogRobust~\cite{zhang2019robust}, group logs by the same predicted log templates and denote them as log events. Then they identify abnormal behaviors based on these log events. On the other hand, some log-based anomaly detection models, such as DeepLog~\cite{deeplog}, pay more attention to the variation of parameters, such as running time or return status code. Unfortunately, most existing work deviates from the original intention of log parsing. Especially, during the evaluation process, the evaluation metrics measure the accuracy of grouping logs but do not explicitly check the extracted templates and parameters. It thus cannot reflect the actual effectiveness of their proposed log parsers. 
The details will be discussed in Section~\ref{subsec:metric}.

\subsection{Insights and Opportunities}
The above three problems motivate us to change our mindset for log parsing.
Through the investigation of multiple sources of logs from public log datasets as well as industrial logs, we uncovered an important characteristic of log data, i.e., even though log contents are very diverse in different systems or services, they generally follow certain common logging practice. Developers may print different log contents, but they have some common logging preferences, which can help others read the log messages easily. For the example in Fig.\ref{fig:example}, developers like to nest a parameter in a bracket ("version (1.0.0)") or place it after a equal sign ("SessionID=30546173"). 
More than that, developers would also conventionally set the status code ("success") or directory path ("/etc/data/") as parameters, rather than fixing them as a part of the template. Actually, there are also many unified and common logging specifications and practices ~\cite{yuan2012characterizing,fu2014developers}, which make logs more informative.

The common logging practice sheds light on the opportunity to capture the common logging behaviours across heterogeneous log sources.
It inspires us that we are able to distinguish the parameters or templates through acquiring conventional logging "syntax".
In order to capture the sharing common logging behaviors across heterogeneous log sources, we need to develop a new log parser, which can learn the common logging practices through understanding the semantic meaning of logs.
After that, the learned model can be applied to parse new sources of log messages.

\section{Approach}
\label{sec:approach}
In this section, we first present the problem formulation in Sec.~\ref{ssec:problem}. We propose a unified log parser based on deep learning technology, named UniParser, to capture the common patterns across heterogeneous log sources indicating parameters or templates.
It contains two major phases, including the offline cross-sources training process and online parsing process, which will be presented in Sec.~\ref{ssec:overview}.
After the overview of UniParser, we will detail the core model architecture design in Sec.~\ref{sec:model_architecture}.
UniParser is composed of three modules, where Token Encoder module and Context Encoder module are utilized for acquiring the semantic patterns from the log token itself and its neighbouring context. Context Similarity module focus more on the commonalities of the learned patterns through contrastive learning. Finally, we will describe the loss function adopted by UniParser in Sec.~\ref{ssec:loss}.

\subsection{Problem Definition}
\label{ssec:problem}

In our proposed model, we transform the log parsing task into the log token classification problem.
Specifically, suppose a raw log message $L$ consists of $n$ tokens after word splitting, denoted as $[t_1, t_2, ... t_n]$. Our proposed log parser is required to predict whether each log token $t_i$ belongs to template part ($y_i=0$) or parameter part ($y_i=1$) based on the learned common logging patterns. All tokens with $y_i=0$ are included in log template and other tokens with $y_i=1$ are put into the parameters list.

\subsection{Overview}
\label{ssec:overview}
\begin{figure}[ht!]
\centering
\includegraphics[width=1.0\linewidth]{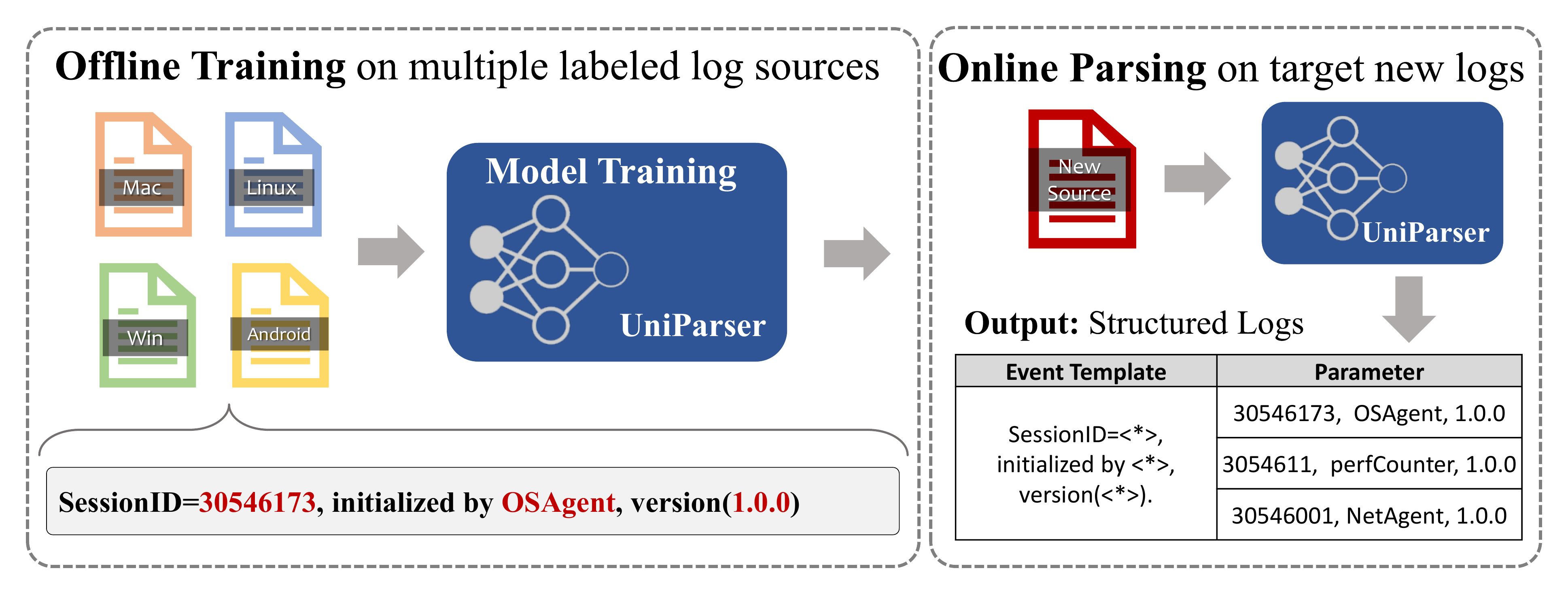}
\caption{An overview of UniParser (red text denotes the parameter part while black text denotes the template part)}
\label{fig:overview}
\end{figure}

In this section, we briefly introduce the overall workflow of UniParser.
As illustrated in Fig.\ref{fig:overview}, there are two phases for the proposed UniParser model, i.e., offline cross-sources training phase and online parsing phase on the target log source. 
During the training phase,  we take labeled log data from multiple sources as a training dataset, which is fed into the UniParser model for training. We expect our model to learn the underlying common patterns indicating templates or parameters instead of the individual and specific log contents.
To achieve this target, we take multiple log sources as the training set and propose a tailored deep learning model, whose architecture and training details will be depicted in Sec.~\ref{sec:model_architecture} and Sec.~\ref{ssec:loss}.

After training across heterogeneous log sources, we can apply our trained model \textit{directly} to the target log source for parsing. 
It will parallelly predict the class (template or parameter) of each log token and then integrate them as log template or parameter list, respectively.
It is worth noting that the training log sources and the testing log sources can be completely mutually exclusive.
Owing to the specially designed deep learning model architecture for capturing the common semantic patterns, UniParser makes generalizing the knowledge learned from other log sources to unknown new log sources possible.
It implies our proposed UniParser does not require any labels in the target log sources to be parsed.

\begin{figure*}[htbp]
\centering
\includegraphics[width=0.9\textwidth]{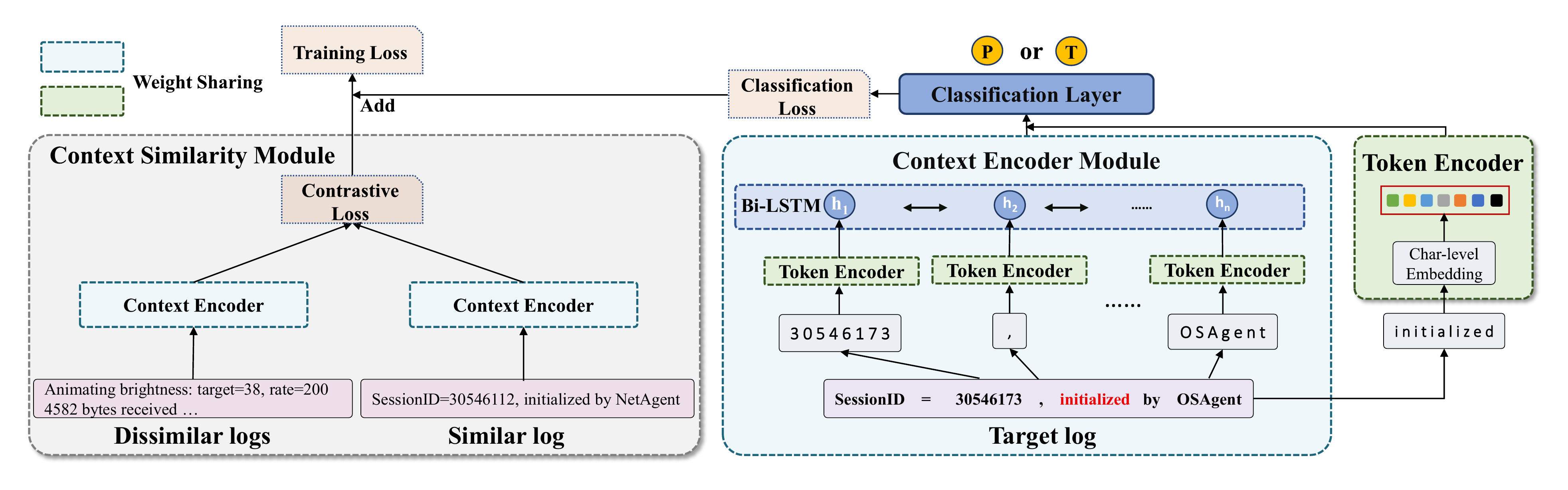}
\caption{The model architecture of UniParser, where the Context Similarity module (left gray part) is used in the offline training phase only. P denotes log parameter token and T denotes log template token.}
\label{fig:architecture}
\end{figure*}

\subsection{Model Architecture}
\label{sec:model_architecture}

\label{ssec:architecture}
In this section, we will elaborate on the model architecture of UniParser.
As shown in Fig.~\ref{fig:architecture}, UniParser consists of three components: Token Encoder module, Context Encoder module and Context Similarity module. 
We will delve into the technical details of all modules in the following subsections.

\subsubsection{Token Encoder module}
\label{ssec:token_encoder}
Given a raw log message, we first conduct word splitting with spaces, tabs, special characters (like equal sign or comma), and so on.
After that, Token Encoder module encodes each split log token into a high-dimensional embedding vector as the model input.
However, diverse and dynamic log words pose a barrier in embedding matrix construction at the token-level.
The endless unseen log tokens would incur the out-of-vocabulary (OOV) problem. 
Developers are permitted to create an infinite variety of variable names, abbreviations or special technical terms, which are far beyond the scale of common English words and will explode the vocabulary.
In addition, some parameters composed of digits (e.g.,"30546173") or some special symbols (e.g.," /etc/data/") are hard to embed as well because they are constantly changing.

To deal with this problem, we alternatively utilize the char-level embedding~\cite{charembedding} to encode log tokens.
The granularity of the character-level is finer than that of the token-level.
The most commonly-used characters, such as lowercase letters, uppercase letters, digits, punctuation and so on (96 characters we used in total), are sufficient to cover the most of tokens formed by their combinations. 
Instead of maintaining a vocabulary of enormous size at the token-level, we build up an embedding layer to encode each character.
Then we sum up all char-level embedding vectors together within in a log token as its encoding vector.
In this way, we can avoid OOV problem and also control the size of the embedding matrix at the appropriate scale.

\subsubsection{Context Encoder module}
\label{ssec:context encoder}
The category (\textit{template or parameter}) of a target log token is not only related to itself but also its neighboring tokens, i.e, \textit{context}.
On the one hand, parameters tend to emerge conjointly with some special characters or symbols.
For the example shown in Fig.\ref{fig:example},   "30546173" is on the heels of an equal sign "=" or "1.0.0" is embraced by a bracket ("(1.0.0)").
On the other hand, the semantics of context is also conducive to indicating the parameters or templates.
A more complicated case is illustrated in line 3 of Fig.\ref{fig:example}.
If we are not so confident in determining the class of log token "/etc/data/" only depending on itself, its context "from" is conducive to classifying it as a parameter because "from" is likely to be followed by a parameter (such as directory address) considering its semantic meaning.

Motivated by the above idea, we utilize Context Encoder module to encode context information. 
Intuitively, not every log token within a log message should be treated as the context of the target log token.
It is because that those log tokens far from the target one tend to be useless for classification.
For example, in line 1 of Fig.\ref{fig:example}, the top token "SessionID" has little relevance to the tail token "(1.0.0)".
Taking that into account, we regard the range from left $k$ to right $k$ tokens around the target token as its context.
In general, $k=3$.
We reuse Token Encoder module to transform each context token into an embedding vector and obtain a sequence of context vectors. 
It is noted that the target token is \textit{masked} during this process.  
The purpose of this operation is that Context Encoder module is expected to independently capture the patterns reflected by context tokens without the aid of the target log token itself.
Otherwise, our model might be biased to the target log token heavily and could not learn the patterns from the context.
The sequence of context vectors is fed into a bidirectional LSTM network to capture the order information.
The hidden states $h_i$ at all time steps of LSTM are aggregated as the final encoding vector of the context.

\subsubsection{Context Similarity module}
Ideally, the embedding vectors produced by Context Encoder module should be close if their corresponding log contexts contain similar patterns.
Nevertheless, the common patterns tend to be manifested with diverse log contents.
Char-level embedding is content-aware and would lead to dispersive encoding vectors since the characters in context might be various.
For the instance in Fig.\ref{fig:example}, when we target to predict the class of token "initialized", its context "by OSAgent" in line 1 and "by perfCounter" in line 2 are very different.
Their encoding vectors may also vary a lot, which is not conducive to making our model capture common patterns in logs.

To overcome this problem, we are required to map the log context under the similar patterns to a tighter vector space, mitigating the effect of the content diversity. We assume that similar log messages are supposed to possess the similar patterns. Thus, we can guide Context Encoder module through the contrast learning between the similar or dissimilar log messages.

To obtain similar log messages, we first cluster the training log data from heterogeneous log sources into several groups. 
The log messages in each group are identical in token length, as well as on the first token.
Log messages in the same group are considered to be similar to each other, while those in different groups are not. After clustering we utilize contrastive learning~\cite{contrastive2021survey} to assist the training of Context Encoder module. For each log message, we randomly select one similar log messages and $|V_d|$ dissimilar ones. The distance between two encoding vectors from similar contexts should be closer. While for the dissimilar ones, their distances should be far apart. This module is optimized towards the above target by constrastive loss~\cite{contrastive2021survey}, which will be introduced in Sec. \ref{ssec:loss}. One interesting point to notice is that the Context Similarity module only serves as an auxiliary module during the training phase and will be disabled when parsing logs online. 
Therefore, Context Similarity module only brings extra time cost during training, while does not slow down the speed of inference.

\subsection{Loss Function}
\label{ssec:loss}
The loss function of our model consists of two parts: token-level classification loss and context-level contrastive loss. 

\paragraph{Token-level classification loss} Collecting the encoding vectors generated from Token Encoder module and Context Encoder module, we concatenate both of them and add the classification layer on top of it to make predictions.
The token-level classification loss can be formulated as:
\begin{equation}
    Loss_{cls} = - \frac{1}{N}(\sum_{i=1}^N[y_i \cdot \log(\hat{y_i}) + (1 - y_i) \cdot (1 - log(\hat{y_i}))])
\end{equation}
As mentioned in Sec.~\ref{ssec:problem}, $y_i$ refers to the label of the target token, and $\hat{y_i}$ is the predicted probability of the token being a parameter. $N$ denotes the total number of tokens. 

\paragraph{Context-level contrastive loss}
The contrastive loss can be formulated as:
\begin{equation}
    Loss_{contrast} = -log\frac{exp(v \cdot v_s)}{ \sum_{v_d \in V_d} exp(v \cdot v_d)}
\end{equation}
where $v$ represents encoding vector for target log context produced by the Context Encoder module, $v_s$ denotes the similar log vector and $v_d$ denotes log vector from the dissimilar set $V_d$, respectively. For each $v$, we randomly select one similar log messages and $|V_d|$ dissimilar log messages ($|V_d|=3$ by default). 
Thanks to the context-level contrastive loss function, UniParser is forced to compress the context encoding vectors from simliar logs much closer than those of dissimilar log messages.

The total loss is the weighted sum of the two losses, which can be formulated as:
\begin{equation}
   Loss = Loss_{cls} + \lambda \cdot Loss_{contrast}
   \label{loss}
\end{equation}
where $\lambda$ is set to 0.01 as default for balancing the magnitude of both loss functions.

\section{Experiment}
\label{sec:experiment}

In order to evaluate the effectiveness and efficiency of UniParser, we conduct extensive experiments.
In this section, we first describe the experiment settings. 
Then, we introduce two evaluation metrics, including group accuracy proposed in \cite{tools}, and another metric proposed in this paper that is better aligned with the real goal of log parsing.
The experiment results of parsing accuracy are presented in Sec.\ref{subsec:sota_comparison}, followed by performance comparison and the component evaluation results.

\subsection{Experimental Settings}
\subsubsection{Datasets}
We conduct experiments based on datasets collected from \textit{LogPai} benchmark \cite{log_parsing_dataset}, which consists of various logs from 16 different systems spanning distributed
systems, supercomputers, operating systems, mobile systems,
server applications, and standalone software. 
Each log message is labeled with a log template as ground truth. 
In our method, we need to split log messages into tokens and separately predict classes of them.
Therefore, we transformed the message-level labels into token-level labels.
Due to some labeling errors in the original version, we also calibrated some labels during the transformation process.

\subsubsection{Implementation Details}

We conduct experiments on a GPU Server equipped with NVIDIA Tesla P100 GPU and CUDA 10.2.
The code is implemented based on PyTorch 1.4.
During the training process, we utilize Adam optimizer and set the initial learning rate as 0.002. We set the batch size as 256 and train the model for 4 epochs. 
During the online parsing phase, we set batch size to 512. 
In Context Encoder module, we set $k=3$. In Context Similarity module, we set $|V_d|=3$.

\subsection{Evaluation Metrics}
\label{subsec:metric}

\textbf{Group Accuracy:} Group Accuracy is proposed in \cite{tools} and has been widely used for evaluating log parsers.
Group Accuracy measures the alignment degree between a set of log messages grouped by identified templates (generated by log parsers) and the corresponding set of log messages belonging to the same true log template. 
However, Group Accuracy prefers properly grouping logs under the same templates together, while the log templates and parameters may not be correctly identified.
For example, the log messages in Fig.\ref{fig:example} are grouped completely correctly by existing log parsers, which indicates the group accuracy is 100$\%$.
However, "/etc/data/" and "success" are misidentified as a part of the template due to their invariance in logs. Therefore, Group Accuracy can not directly reflect whether messages are correctly parsed by log parsers .

\begin{figure}[ht!]
\centering
\includegraphics[width=1.0\linewidth]{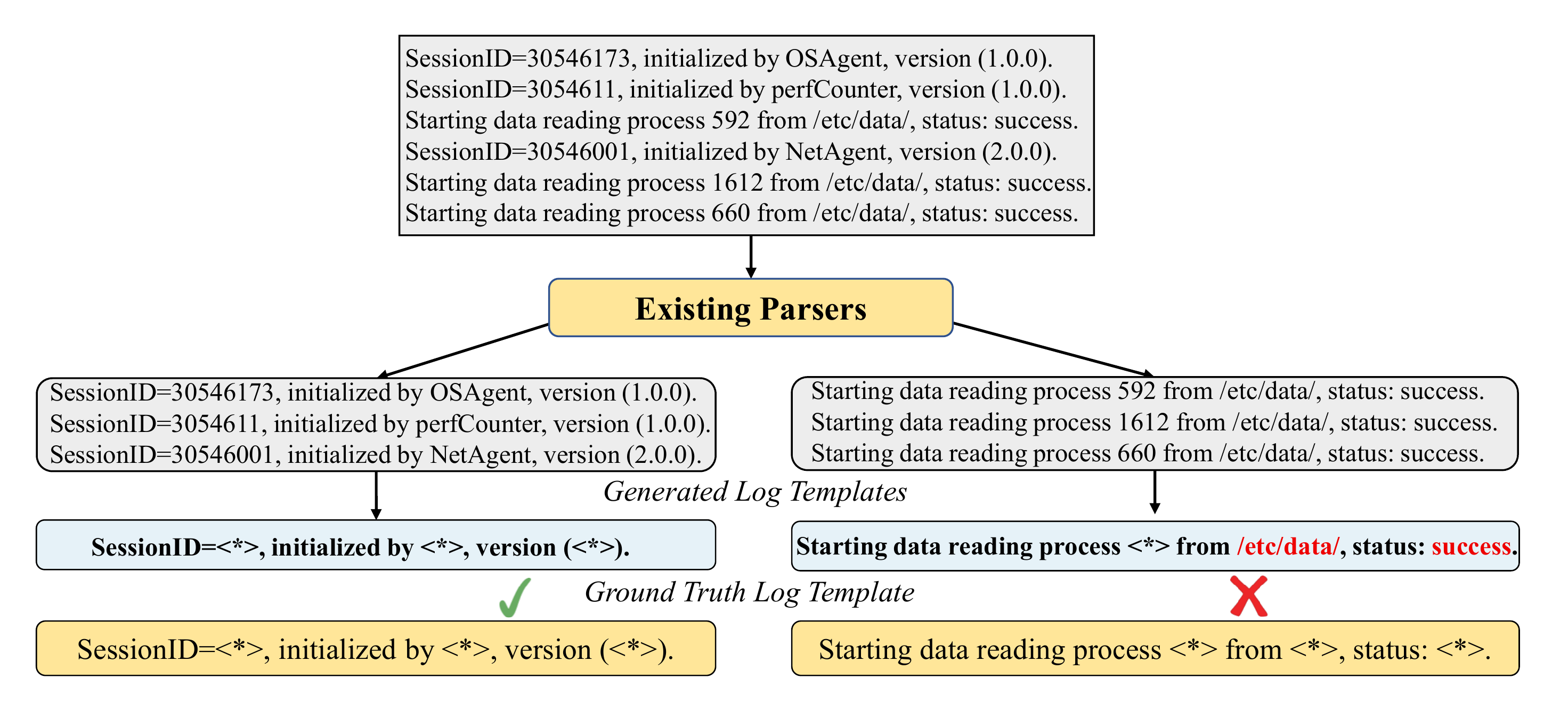}
\caption{Existing log parsers misidentify parameters ("etc/data/" and "success") but group logs under the same template correctly}
\label{fig:weak}
\end{figure}

\textbf{Message-Level Accuracy (MLA):} we propose Message-Level Accuracy to overcome the shortcomings of Group Accuracy, where a log message is considered correctly parsed if and only if every token of the message is correctly identified as template or parameter. 
Obviously, this metric is much stricter than Group Accuracy since any incorrect log token identification will lead to the wrong parsing result for the whole log message. 
For example, MLA in Fig.\ref{fig:weak} is $\frac{3}{6}$ since only logs in left group (3 log messages) are parsed completely correctly.

\subsection{Parsing Accuracy Evaluation}
\label{subsec:sota_comparison}
We compare our proposed model with four state-of-the-art methods (including Drain~\cite{drain}, AEL~\cite{jiang2008abstracting}, LenMa~\cite{LenMa}, and LFA~\cite{LFA}) on all 16 log datasets. These log parsers cover most types of parsing techniques (see Sec.\ref{ssec:related_work}) and achieve good results on public datasets.
For each dataset to be tested, we conduct cross-source training based on the other 15 datasets and applied the trained model to the target dataset.
Since Group Accuracy measures the overlap between logs sets under the ground truth templates and the predicted templates given by parsers, we thus needed to group logs under the similar predicted log templates together for evaluation.
The results in terms of two metrics (Group Accuracy, Message-Level Accuracy) are shown in Table \ref{table:group} and Table \ref{table:template}, respectively.

From the results, we can see that our model outperforms these existing methods on almost all datasets with respect to the two metrics. 
Specifically, compared with the powerful existing log parser Drain~\cite{drain}, UniParser exceeds it by 11.9\% on group accuracy and 40.8\% on MLA on average.
In addition, we also note that traditional log parsers such as Drain achieves high group accuracy (0.867) but a very low MLA (0.377). 
It confirms that existing methods focus more on grouping logs under the same template, but ignore the identification of templates and parameters. Different from the related work, UniParser is capable of learning semantic patterns and predicting the categories token by token. 
Therefore, it achieves a higher score on MLA (0.785), demonstrating the effectiveness of our method. 

One notable issue is that UniParser does not outperform existing methods on Proxifier dataset. 
After investigation, we found that there exists a large gap between Proxifier and other datasets,  which decreases the performance of UniParser.
We will discuss this problem in Sec.\ref{sec:discussion} and show how to improve the accuracy of UniParser via a small amount labeled data for fine-tuning.

\begin{table}[ht!]
\centering
\caption{Comparison with the state-of-the-art log parsers on Group Accuracy}
\label{table:group}
\begin{tabular}{c|ccccc}
\toprule
Method & Drain & AEL & LenMa & LFA & UniParser \\
\midrule
HDFS      &0.998  & 0.998 & 0.998 & 0.885 & \textbf{1.000} \\ 
Hadoop    &0.948  & 0.538  & 0.885 & 0.900 & \textbf{1.000}\\
Spark     &0.920 & 0.905 & 0.884 & 0.994 & \textbf{1.000}\\
ZooKeeper &0.967  & 0.921 & 0.841 & 0.839 & \textbf{0.995}\\
OpenStack &0.733   & 0.758 & 0.743 & 0.200 & \textbf{1.000}\\
BGL       &0.963   & 0.758 & 0.690 & 0.854 &\textbf{ 0.997}\\
HPC       &0.887  & 0.903 & 0.830 & 0.817& \textbf{0.966}\\
Thunderbird &0.955 & 0.941 & 0.943 & 0.649 & \textbf{0.990}\\
Windows   &0.997  & 0.690 & 0.566 & 0.588 & \textbf{1.000}\\
Linux     &0.690 & 0.673 & 0.701 & 0.279 & \textbf{0.878}\\
Mac       &0.787  & 0.764 & 0.698 & 0.599 & \textbf{0.997}\\
Android   &0.911  & 0.712 & 0.880 & 0.616 & \textbf{0.973}\\
HealthApp &0.780  & 0.822 & 0.174 & 0.549 &\textbf{1.000}\\
Apache    &1.000  & 1.000  & 1.000& 1.000 & \textbf{1.000}\\
OpenSSH   &0.788  & 0.538 & 0.925 & 0.501 & \textbf{1.000}\\
Proxifier &0.527  & 0.518 & 0.508 & 0.0026 & \textbf{0.976}\\ \midrule
Average & 0.867	& 0.777 & 0.767 & 0.642 & \textbf{0.986}	\\
\bottomrule
\end{tabular}
\end{table}

\begin{table}[ht!]
\centering
 \caption{Comparison with the state-of-the-art log parsers on MLA}
\label{table:template}
\begin{tabular}{c|ccccc}
\toprule 
Method & Drain & AEL & LenMa & LFA & UniParser \\
\midrule 
HDFS      &0.567  & 0.568 & 0.123 & 0.156 & \textbf{1.000} \\ 
Hadoop    &0.530  & 0.526  & 0.079 & 0.499 & \textbf{0.866}\\ 
Spark     &0.384 & 0.373 & 0.006 & 0.382 & \textbf{0.972}\\
ZooKeeper &0.792  & 0.748 & 0.677 & 0.340 & \textbf{0.99}\textbf{2}\\
OpenStack &0.019   & 0.021 & 0.018 & 0.008 & \textbf{0.459}\\
BGL       &0.341   & 0.341 & 0.082 & 0.230 & \textbf{0.811}\\
HPC       &0.701  & 0.725 & 0.632 & 0.674& \textbf{0.974}\\
Thunderbird &0.059 & 0.048 & 0.038 & 0.026 & \textbf{0.542}\\
Windows   &0.158  & 0.153 & 0.152 & 0.142 & \textbf{0.691}\\
Linux     &0.169 & 0.164 & 0.107 & 0.023 & \textbf{0.854}\\
Mac       &0.176  & 0.148 & 0.094 & 0.082 & \textbf{0.584}\\
Android   &0.431  & 0.350 & 0.430 & 0.299 & \textbf{0.838}\\
HealthApp &0.295  & 0.295 & 0.174 & 0.285 &\textbf{ 0.985}\\
Apache    &0.694  & 0.690  & 0.634 & 0.688 & \textbf{0.994}\\
OpenSSH   &0.507  & 0.247 & 0.133 & 0.164 & \textbf{0.626}\\
Proxifier &0.203  & 0.195 & 0.017 & \textbf{0.478} & 0.369\\ \midrule
Average & 0.377 & 0.350 & 0.212 & 0.280 & \textbf{0.785} \\
\bottomrule
\end{tabular}
\end{table}

\begin{figure*}[t]
\centering
\begin{minipage}[t]{.30\textwidth}
\includegraphics[width=\linewidth]{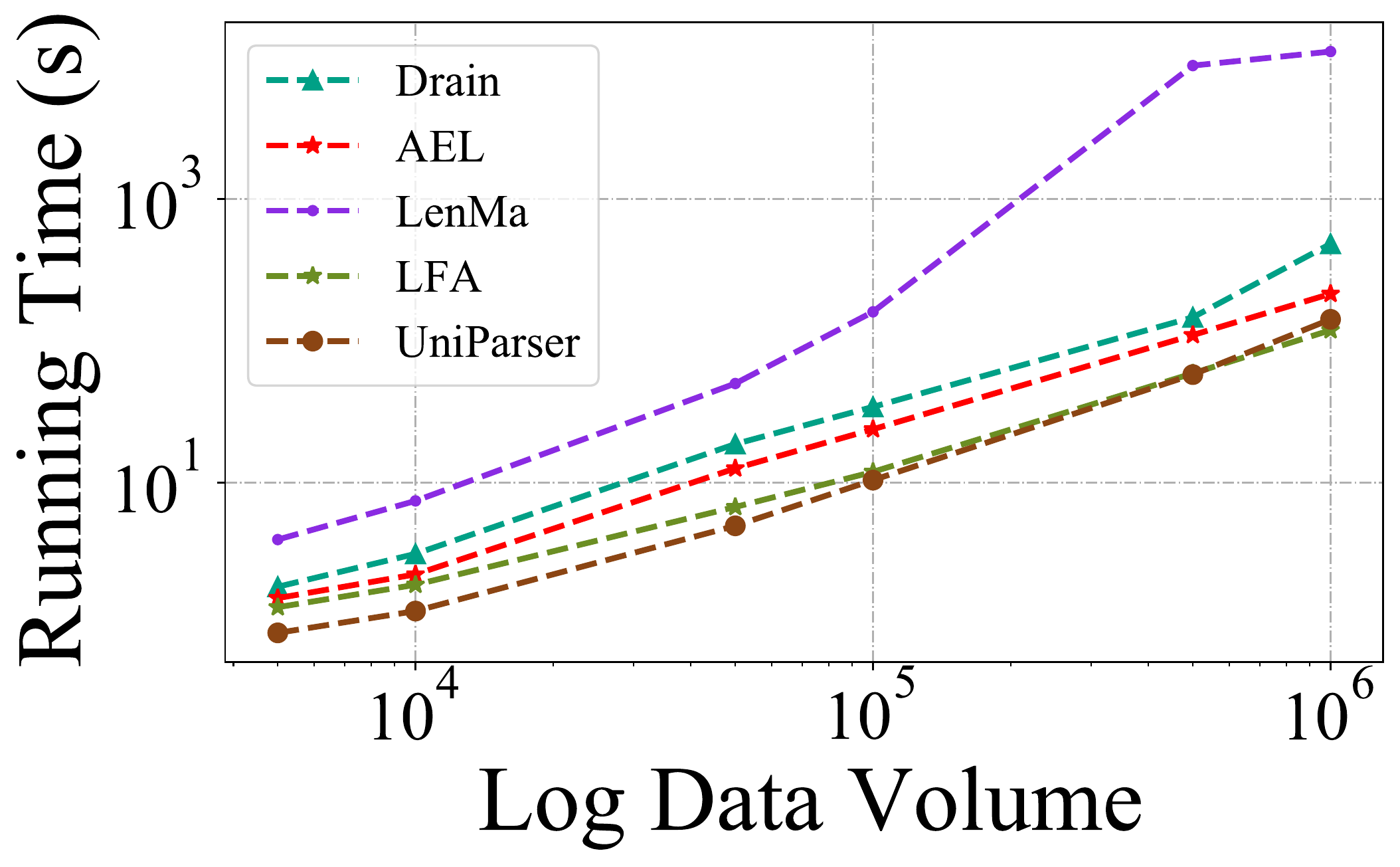}
\caption{Running time of different log parsers under different volume.}
\label{fig:performance}
\end{minipage}\qquad
\begin{minipage}[t]{.30\textwidth}
\includegraphics[width=\linewidth]{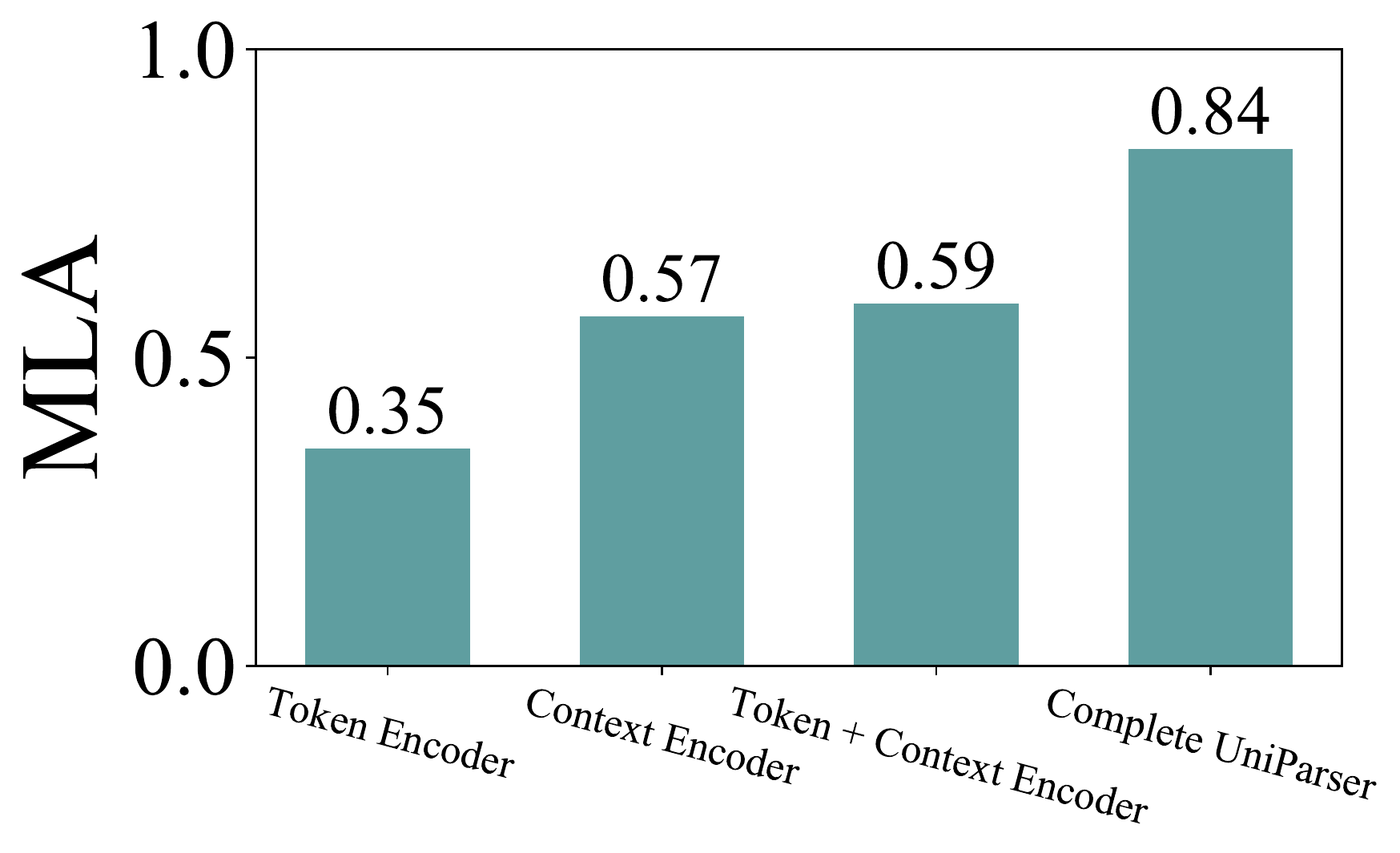}
\caption{Ablation study results.}

\label{fig:ablation_study}
\end{minipage}\qquad
\begin{minipage}[t]{.30\textwidth}
\includegraphics[width=\linewidth]{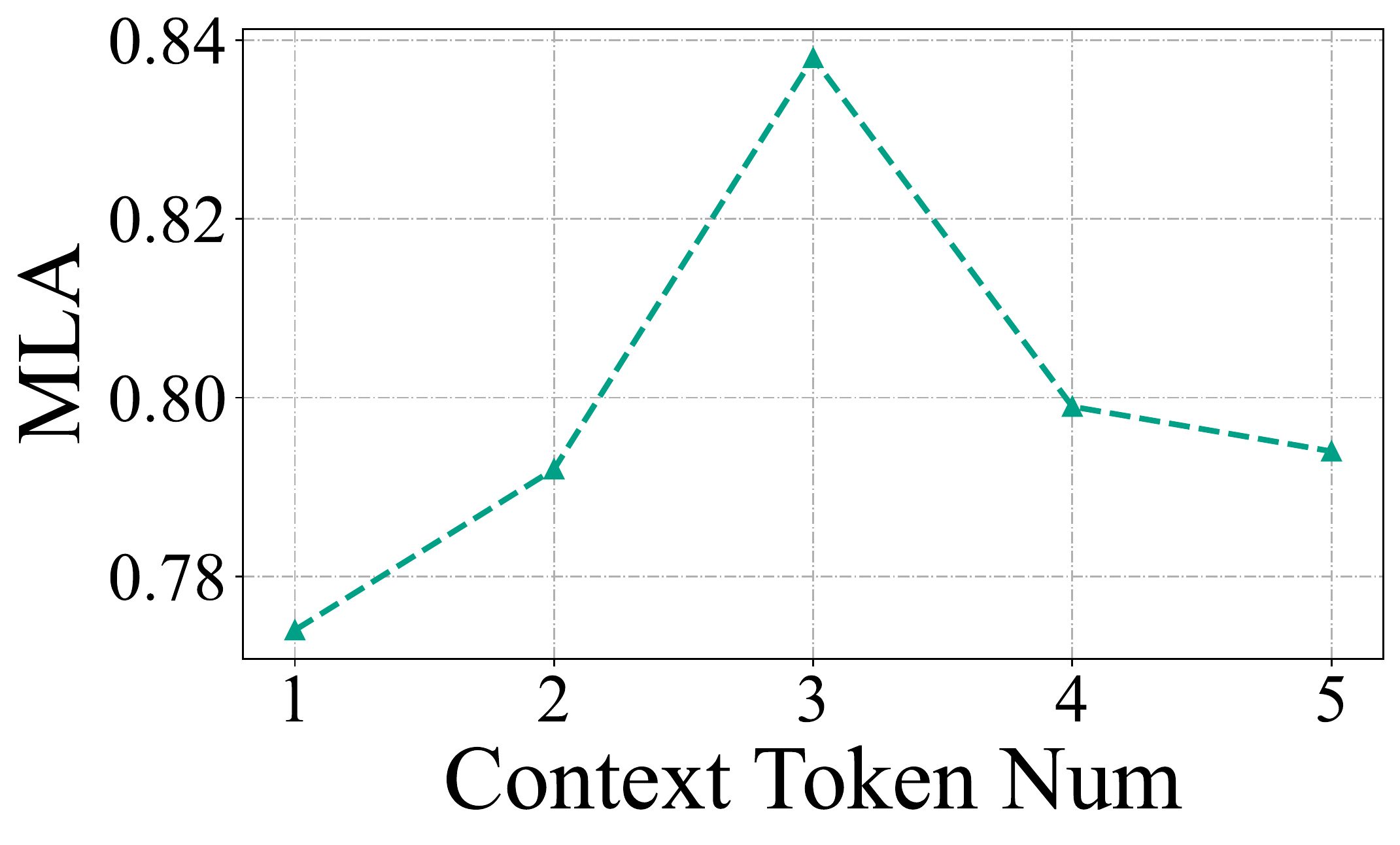}
\caption{Parameter exploration of Context Encoder module}
\label{fig:params_test}
\end{minipage}
\vspace{-6pt}
\end{figure*}

\subsection{Runtime Performance Evaluation}
Besides parsing accuracy, performance is another critical metric for log parsers. 
Therefore, we also need to compare the running time of UniParser with other log parsers under different volumes of log data. 
It is worth noting that we also include the time cost of loading data for our proposed UniParser. 
The results are shown in Fig.\ref{fig:performance} (X and Y axis both are in log-scale).
From the results, we can see the running time of UniParser increases slowly with the log scale expansion.
Moreover, with the parallel prediction of log tokens based on GPU acceleration, our model performs faster than other traditional parsers.
Even at the scale of one million log messages, UniParser took about 140 seconds, which is only about half of the time spent by AEL (214s) and around a quarter of the time spent by Drain (482s).

\subsection{Component Evaluation}
In this section, we evaluate the effectiveness of the major components and parameters. 
The experiments are performed based on Android log dataset as it is enriched with hundreds of log templates and complicated patterns, which is representative among 16 log datasets. As mentioned in Sec.\ref{subsec:metric}, Group Accuracy can not properly measure the performance of log parsers, we only focus on MLA metric in this section.

\subsubsection{Ablation Study}
\label{ssec:ablation}
Firstly, we explore the effectiveness of each component on our model, and the results are shown in Fig.\ref{fig:ablation_study}.  
UniParser exhibits worse accuracy with only Token Encoder module (0.352 MLA), while achieves a rapid boosting with Context Encoder module (0.587 MLA) added. 
This comparison indicates that Context Encoder module learns common patterns from context, which facilities the target token classification. 
In addition, Context Similarity module significantly improves the accuracy of UniParser (0.838 MLA), which demonstrates its usefulness for making Context Encoder module encodes feature vector more precisely.

\subsubsection{Parameter Analysis}
As shown in Fig.\ref{fig:params_test}, we explored the effect of $k$ in Context Encoder module, which denotes the number of tokens around the target token (as mentioned in Sec.\ref{ssec:context encoder}).
We adjusted the value of $k$ from 1 to 5 and observe the changes in MLA.
From the results, we found that MLA is relatively stable in the whole range (from 0.78 $\sim$ 0.84). It increases gradually when $k$ in [1, 3] but drops if $k$ continues increasing. This implies that context tokens indeed help to capture semantic patterns. On the contrary, tokens far away from target token may not provide effective context information, and could bring noise to the model and decrease the accuracy of UniParser. 

\subsubsection{Comparison with sequence labeling model}
In addition, we compared our model with Bidirectional LSTM-CNNs-CRF \cite{ma2016end}, a commonly-used sequence labeling model which can be utilized to classify log tokens as well. 
Bi-LSTM-CNNs-CRF first utilizes CNN to encode each word into a vector representation, then word-level representations are fed into Bi-LSTM to obtain context information. Lastly, on top of Bi-LSTM, a sequential CRF jointly decode labels for the whole sentence. On the one hand, the results in Table~\ref{table:crf} indicate that deep learning based log parsers benefit from the supervised labels on token-level and indeed perform better than previous unsupervised methods (both models outperform other existing log parsers such as Drain on Android dataset); 
On the other hand, our UniParser outperforms Bidirectional LSTM-CNNs-CRF, 
which demonstrates the effectiveness of our proposed model.
 
\begin{table}[H]
\centering
\caption{Comparison with Bi-LSTM-CRF}
\label{table:crf}
\begin{tabular}{c|ccc}
\toprule
Method & Drain & Bi-LSTM-CNNs-CRF & UniParser\\
\midrule
MLA & 0.431 & 0.555 &\textbf{0.838} \\
\bottomrule
\end{tabular}
\end{table}

\section{Discussion}
\label{sec:discussion}

\paragraph{Model Fine-tuning:}
Patterns may differ between the training and the target log sources. To tackle this problem, we devise an online feedback mechanism to fine-tune the UniParser model under the target log data.
Engineers only need to inspect a few lines of parsed structured logs and calibrate the results (templates or parameters) according to their domain knowledge.
Then these labeled logs are fed into UniParser to make the model fast adapt to the new log sources.
We applied fine-tuning to our model on Proxifier dataset, where UniParser performs worst. The result is listed in Table~\ref{tab:fine_tune}.
It can be seen that after fine-tuning with tens of labeled logs, the UniParser model can achieve a rapid boosting of parsing accuracy.

\begin{table}[ht!]
\centering
\caption{Model fine tuning with different numbers of labeled logs}
\label{tab:fine_tune}
\begin{tabular}{c|ccc}
\toprule
\#Samples & 0 & 20 & 40 \\ 
\midrule
MLA & 0.369 & 0.507& 0.893 \\ 
\bottomrule
\end{tabular}
\end{table}

\paragraph{Labeling Effort}
Our proposed method relies on the labeled log data, which means some labeling effort is required for training.
The labeled log datasets are used in the offline supervised learning and can be labeled once for all.
From the experiment results shown in Sec.\ref{subsec:sota_comparison}, our model is capable of learning the common logging patterns from heterogeneous log sources, and can be applied to most of the new log sources directly without extra labeling effort.
Even for the logs with distinctive patterns such as  those in the Proxifier dataset, a small amount of fine-tuning is sufficient.

\paragraph{Token Splitting}
As introduced in Sec.\ref{ssec:problem}, UniParser transforms the log parsing task into the log token classification problem.
We split tokens in raw log messages according to some special symbols such as spaces, tabs and comma (see Sec.\ref{ssec:token_encoder}).
Nevertheless, some log tokens are difficult to be split up due to the complexity of the log messages. 
For example, in log message "Process A done this 1 time(s)", "(s)" in the bracket should not be separated from the preceding token "time" and is not a parameter.
In our future work, we will design effective mechanisms to handle rare cases like this.
\section{Application in Practice}
So far, our proposed UniParser has already successfully applied to multiple log analysis scenarios in Microsoft 365 and Azure Cloud, including safe deployment guarding,  performance issue diagnosis, log-based anomaly detection and so on.
Compared with the existing log parsers, UniParser achieves much higher parsing accuracy and exhibits more robustness on complicated industrial log data, which greatly improves the performance of various log-based downstream tasks.

\section{Conclusion}
\label{sec:conclusion}
Log parsing is the foundation of automated log analysis, which is responsible for transforming the semi-structured raw logs into a structured format. In this paper, we propose a novel deep learning based model named UniParser to capture the common logging behaviours from heterogeneous log data.
UniParser utilizes a Token Encoder module and a Context Encoder module to capture the patterns of templates and parameters.
A Context Similarity module is specially designed to focus on the commonalities of learned patterns.
We have evaluated UniParser on public log datasets and the results show that UniParser outperforms the state-of-the-art parsers by 12\% on Group Accuracy and about 40\% on Message-Level Accuracy. 

{\setcounter{secnumdepth}{-2}}
\section{ACKNOWLEDGMENTS}
We would like to thank Murali Chintalapati, Jim Kleewein, Andrew Zhou, Zhangwei Xu, Thomas Moscibroda, Victor Rühle, Chetan Bansal, Lingling Zheng, Marcus Fontoura and Silvia Yu for
their great support and sponsorship.

\balance
\bibliographystyle{ACM-Reference-Format}
\bibliography{sample-base}

\end{document}